\documentclass[12pt]{article}
\usepackage{graphicx}
\usepackage{natbib}
\usepackage{amssymb}

\begin{document}

\title{\bf Chaotic zones around gravitating binaries}

\author{Ivan~I.~Shevchenko\/\thanks{E-mail:~iis@gao.spb.ru} \\
Pulkovo Observatory of the Russian Academy of Sciences \\
Pulkovskoje ave.~65, Saint Petersburg 196140, Russia}
\date{}

\maketitle

\begin{center}
Abstract
\end{center}

\noindent The extent of the continuous zone of chaotic orbits of a
small-mass tertiary around a system of two gravitationally bound
primaries of comparable masses (a binary star, a binary black
hole, a binary asteroid, etc.) is estimated analytically, as a
function of the tertiary's orbital eccentricity. The separatrix
map theory is used to demonstrate that the central continuous
chaos zone emerges (above a threshold in the primaries' mass
ratio) due to overlapping of the orbital resonances corresponding
to the integer ratios $p$:1 between the tertiary and the central
binary periods. In this zone, the unlimited chaotic orbital
diffusion of the tertiary takes place, up to its ejection from the
system. The primaries' mass ratio, above which such a chaotic zone
is universally present at all initial eccentricities of the
tertiary, is estimated. The diversity of the observed orbital
configurations of biplanetary and circumbinary exosystems is shown
to be in accord with the existence of the primaries' mass
parameter threshold.

\bigskip


\newpage

\section{Introduction}

Zones of orbital instability are known to exist around binary
stars. In particular, such zones are known to be present in
circumbinary protoplanetary disks. The latter may contain
planetesimals, dust, and gas. Gas is present on the initial stages
of the disk evolution; it dissipates later on. Many recent
numerical simulations show that, irrespective of the gas content,
a free-from-matter central cavity always forms around a system of
two gravitationally bound bodies of comparable masses; see
\citep{MN04,PN07,PN08,M12,PLTB12} for planetesimal disks with and
without gas.

The existence and possible characteristics of central cavities in
gaseous circumbinary disks were considered theoretically as early
as 1994 by \cite{AL94,AL96}, in view of the observational data on
the disks around such stars as GW~Ori. Artymowicz and Lubow
outlined the role of Lindblad resonances in the cavities
formation, the cavity size increasing with decreasing the disk
viscosity (see, e.g., Figure~4 in \citealt{AL96}).

In a quite separate field of study, namely, the dynamical studies
of triple stars, it is well known that the stability of
hierarchical triple stars is mostly determined by the pericentric
distance of the tertiary~\citep{M08,V08,STO12,STO13}: if this
distance is below a critical one, the system is unstable. A number
of heuristic semi-analytical criteria for the critical pericentric
distance were proposed (see reviews in \citealt{M08,V08}). Most of
these criteria do not appeal to resonant phenomena, with a few
exceptions. \cite{MW06} explored the overlapping of subresonances
of mean motion resonances, which is responsible for defining the
local stability borders in the phase space of motion of a planet
orbiting a binary, and derived expressions for the subresonance
borders. In the general three-body problem, the role of overlap of
orbital resonances for the stability of triples was considered by
\cite{Ma08}, who developed an algorithm for constructing the
analytical stability border as an envelope of a set of local
resonance borders. For the hierarchical triples, Mardling revealed
the major role (for the formation of the global stability border)
of overlapping of the orbital resonances, corresponding to the
integer ratios $p$:1 between the tertiary and the binary periods.

On the other hand, numerical-experimental criteria are of great
value (due to their high comparative accuracy) and are mostly used
nowadays in studies of the stability of planets in binary systems,
especially, in the dynamical studies of circumbinary planets.
\cite{HW99} derived a numerical-experimental criterion for the
radius of the central chaotic zone, in the framework of the planar
restricted three-body problem. The observational data on the
recently discovered exoplanets orbiting around main-sequence
binaries (Kepler-16b, 34b, 35b, and others), combined with
estimates based on the Holman--Wiegert criterion, shows that these
circumbinary planets move in orbits encircling the central chaotic
zone, and usually these orbits are close to its
boundary~\citep{D11,W12,W14,PS13}.

Nowadays, the circumbinary dynamics is studied mostly in the
exoplanetary context. However, the presence of a central chaotic
zone is, of course, universal for the dynamics around any
gravitating binary with comparable component masses (a binary
star, a binary black hole, a binary asteroid, etc.).

In this paper, the extent of the continuous zone of chaotic orbits
of a small-mass tertiary around a system of two gravitationally
bound bodies of comparable masses is estimated strictly
analytically (i.e., without introducing any empirical or heuristic
factors). The binary's mass ratio, above which such a chaotic zone
is universally present, is also estimated. Chirikov's resonance
overlap criterion~\citep{C59,C79} and the general separatrix map
theory~\citep{S07IAU,S10,S11} are used. The analysis is performed
in the framework of the planar restricted circular three-body
problem, i.e., the eccentricity of the central binary is zero, and
the orbit of the test zero-mass particle is coplanar with that of
the binary.

The paper is organized as follows. In Section~\ref{sec_crit}, we
briefly review the existing relevant criteria for the orbital
stability of three-body systems, then we consider the separatrix
map theory and, based on it, derive a new, strictly analytical
criterion for the appearance of a circumbinary zone of the
tertiary's chaotic orbits. In Section~\ref{sec_size}, using the
derived criterion, we estimate the typical size of the
circumbinary chaotic zone. It is demonstrated that the central
continuous chaos zone emerges due to overlapping of the orbital
resonances corresponding to the integer ratios $p$:1 between the
tertiary and the central binary periods. In Section~\ref{sec_thr},
we consider a mass parameter threshold, above which the central
chaotic zone is universally present at all initial eccentricities
of the tertiary. In Section~\ref{sec_diag}, we analyze the
theoretical chaos border in a ``pericentric
distance--eccentricity'' stability diagram for exosystem Kepler-16
(the prototype of circumbinary exosystems), and inspect how well
the theoretical border fits the numerical-experimental one. In
Section~\ref{sec_thr_obs}, we discuss the relevance of the mass
parameter threshold to the diversity of the observed orbital
configurations of biplanetary and circumbinary exosystems.
Section~\ref{sec_concl} is devoted to our conclusions.

\section{An analytical criterion based on the Kepler map}
\label{sec_crit}

First of all, let us recall the case in which the masses of the
binary components are not comparable, i.e., one of them is much
less than the other one. Then, as shown by \cite{W80}, a narrow
annular chaotic band exists, surrounding the orbit of the
secondary. The inner half-band forms due to accumulation of the
first-order resonances $(p+1)$:$p$ between the orbital periods of
the secondary and the tertiary,\footnote{The bodies are called the
primary, the secondary, and the tertiary in accord with the
hierarchy of their masses. Thus, the primary is the main
gravitating body, the secondary is the perturber, and the tertiary
is the test particle.} and the outer one forms due to the
accumulation of the resonances $p$:$(p+1)$ between the same
periods. With increasing $p$, the resonances group more densely,
and, at some threshold value of $p$, they start to overlap because
the resonance width does not shrink fast enough with $p$.
Chirikov's resonance overlap criterion states that when resonances
overlap, global chaos emerges~\citep{C59,C79,LL92}. On these
grounds, \cite{W80} showed, in the framework of the planar
circular restricted three-body problem, that in the case of small
eccentricity ($e \lesssim 0.15$) of the particle's orbit, the
value of $p$, critical for the overlap of the first-order
resonances, is given by

\begin{equation}
p_\mathrm{overlap} \approx 0.51 \mu^{-2/7},
\label{pcrit}
\end{equation}

\noindent where $\mu = m_2 / (m_1 + m_2)$ is the mass parameter of
the binary; $\mu \ll 1$. Let $a$ and $a'$ be the semimajor axes of
the particle's and perturber orbits, respectively; then, using
Equation~(\ref{pcrit}) and Kepler's third law, it is
straightforward to find the interval $\Delta a_\mathrm{overlap} =
| a - a' |$, where the $(p+1)$:$p$ resonances overlap:

\begin{equation}
\Delta a_\mathrm{overlap} \approx 1.3 \mu^{2/7} a'
\label{em_27}
\end{equation}

\noindent \citep{DQT89,MD99}. The particles with $a \in a' \pm
\Delta a_\mathrm{overlap}$ move chaotically. Thus, the radial
extent of the instability neighborhood of the perturber's orbit is
given by the ``$\mu^{2/7}$ law''.

\cite{MD99} checked the validity of this analytical result in
direct numerical integrations of orbits. It turned out that in a
studied interval of $\mu$ (which was rather broad: $10^{-9} \le
\mu \le 10^{-3}$) the numerically derived power-law index is equal
to $0.286$, practically coinciding with the theoretical prediction
$2/7$, but the coefficient $1.3$ is in fact $\sim 30\%$ greater:

\begin{equation}
\Delta a_\mathrm{overlap} = 1.57 \mu^{0.286} a'
\label{actem_27}
\end{equation}

\noindent (see Figure~9.23 in \citealt{MD99}). This means that the
real chaotic annular zone is about $30\%$ broader, regardless of
the $\mu$ value.

Formally extrapolating Equation~(\ref{actem_27}) to $\mu = 1/2$
(the equal-mass case), one finds $\Delta a_\mathrm{overlap} /
a^\prime \approx 1.29$, i.e., the zone inside the binary is
expected to be continuously chaotic, and the outer border of this
continuous chaotic zone is expected to be situated at $\approx
(0.5 + 1.29) a' \approx 1.8 a'$ from the barycenter (the binary's
center of mass). We shall see that this severely underestimates
the actual size (in fact, the actual size $\approx 2.8 a'$). The
reason for the discrepancy is not just a numerical uncertainty of
extrapolation. It is deeply physical: as we shall see, the class
of resonances responsible for the formation of the central chaotic
zone is totally different from that responsible for the formation
of the annular chaotic band.

As massive numerical data show (see \citealt{HW99,V08}), when the
binary's components are comparable in mass, the central continuous
chaotic zone indeed exists. The zone size depends, in particular,
on the tertiaries' initial eccentricities \citep{V08}; if they are
zero, then the size can be estimated (in the planar restricted
three-body problem) using the Holman--Wiegert
numerical-experimental criterion \citep{HW99}: the radius
$a_\mathrm{cr}$ of the instability zone for the initially circular
prograde circumbinary orbits is given by the smooth fitting
function

\begin{equation}
a_\mathrm{cr}/a_\mathrm{b} = 1.60 + 5.10 e_\mathrm{b} - 2.22
e_\mathrm{b}^2 + 4.12 \mu - 4.27  e_\mathrm{b} \mu - 5.09 \mu^2 +
4.61 e_\mathrm{b}^2 \mu^2 ,
\label{HWcrit}
\end{equation}

\noindent where $\mu = m_2 / (m_1 + m_2)$, and $a_\mathrm{b}$ and
$e_\mathrm{b}$ are the semimajor axis and eccentricity of the
binary. Thus, the Holman--Wiegert criterion,
Equation~(\ref{HWcrit}), utilizes a polynomial fit over numerical
data in $a_\mathrm{b}$ and $e_\mathrm{b}$ for the description of
the global instability border location. However, note that this
border is in fact fractal \citep{PS12,PS13}; an example of its
``ragged'' appearance will be given below.

In what follows, we analytically explore the phenomenon of the
circumbinary chaos zone. Similarly to the above-mentioned
accumulation of the first-order resonances $(p+1)$:$p$ in the
close-to-coorbital motion, there exists an accumulation of
resonances $p$:1 in the circumbinary highly eccentric motion,
close to the parabolic one.

To describe this accumulation of resonances, we use the ``Kepler
map'' theory, initiated by \cite{P86}, \cite{PB88} and
\cite{CV86}. They found that if one writes down the expression for
the tertiary's energy $E$ increment together with the expression
for the time increment between two consecutive pericenter passages
by the tertiary (this time increment is directly proportional to
the increment of perturber's phase angle $g$, and is expressed
through the energy via Kepler's third law), one obtains a
two-dimensional area-preserving map, the so-called Kepler map:

\begin{eqnarray}
E_{i+1} &=& E_{i} + W \sin g_{i} , \nonumber \\
g_{i+1} &=& g_{i} + 2 \pi \vert 2 E_{i+1} \vert^{-3/2} ,
\label{kmp_gm2}
\end{eqnarray}

\noindent where the subscript $i$ enumerates the pericenter
passages. (On the general theory of area-preserving maps, see
\citealt{M92}.) As derived in \cite{S11}, in the case of prograde
(with respect to the binary) tertiary orbits, the coefficient $W$
is given by

\begin{equation}
W \simeq 2^{1/4} \pi^{1/2} \mu q^{-1/4} \exp \left( -
\frac{2^{3/2} q^{3/2}}{3} \right) ,
\label{kmp_Wqasy}
\end{equation}

\noindent if $\mu \ll 1/2$ and $q = a(1-e) > 1$ ($q$ is the
tertiary's pericentric distance). Hereafter we set the
gravitational constant $G=1$, the sum of the masses of the
binary's components $m_1 + m_2 = 1$, and the binary's semimajor
axis $a_\mathrm{b}=1$ (then the binary's period is equal to $2
\pi$).

By means of substitution $E = W y$, $g = x$, map~(\ref{kmp_gm2})
is reducible to

\begin{eqnarray}
     y_{i+1} &=& y_i + \sin x_i, \nonumber \\
     x_{i+1} &=& x_i + \lambda \vert y_{i+1} \vert^{-3/2} ,
\label{kmp_gm1}
\end{eqnarray}

\noindent where

\begin{equation}
\lambda = 2^{-1/2} \pi W^{-3/2} .
\label{kmp_la}
\end{equation}

\noindent Thus, the Kepler map is parameterized by the single
parameter $\lambda$~\citep{S10}. Both maps (\ref{kmp_gm2}) and
(\ref{kmp_gm1}) depend on a single parameter, be it $\lambda$ or
$W$, but the advantage of Equations~(\ref{kmp_gm1}) over
(\ref{kmp_gm2}) is that the $\lambda$ parameter in
Equations~(\ref{kmp_gm1}) is an analog of the adiabaticity
parameter in the case of the classical separatrix map (derived in
\citealt{C79}); therefore, by its value, one can judge whether
chaos is adiabatic or not. (The term ``adiabatic chaos'' concerns
the conservation of an adiabatic invariant; at low values of
$\lambda$, it is conserved on long time intervals between
crossings of the separatrix, see \citealt{CV00,S08}.)

The Kepler map is an example of a general separatrix
map~\citep{S10}, the separatrix (the $y=0$ line) separating the
bound and unbound states of the particle's motion. At $q \gg 1$,
one has $W \ll 1$ (see Equation~(\ref{kmp_Wqasy})), therefore
$\lambda \gg 1$. This means that chaos in the motion of particles
is not adiabatic~\citep{S07IAU}, and therefore the Kepler map can
be locally approximated by the standard map with good
accuracy~\citep{S10,S11}.

The Kepler map is derived in the assumption that the tertiary's
pericentric distance $q$ is constant. This can be justified quite
easily, using the classical Tisserand relation (based on the
Jacobi constant formalism). In the planar circular restricted
three-body problem, the Tisserand relation is given by

\begin{equation}
\frac{1}{a} + 2  \left[ (1-e^2) a \right]^{1/2} \approx
\mathrm{const} , \label{Tiss1}
\end{equation}

\noindent where $a$ is the semimajor axis of the particle,
measured in the units of the perturber's semimajor axis $a'$, $e$
is the particle's eccentricity (see, e.g., \citealt{MD99}). If $a
\gg a'$, and $e \sim 1$, one has

\begin{equation}
\frac{1}{a} + 2  \left[ (1+e) q \right]^{1/2} \approx 2^{3/2}
q^{1/2} \approx \mathrm{const} . \label{Tiss2}
\end{equation}

\noindent Thus, $q$ is approximately conserved.

\cite{P86} used the Kepler map theory to show that the energy
width of a one-sided chaotic band in the vicinity of the perturbed
parabolic orbit scales as the power $2/5$ of the mass parameter:

\begin{equation}
\Delta E_\mathrm{cr} = \left| E_\mathrm{cr} \right| = -
E_\mathrm{cr} \propto \mu^{2/5} ,
\label{Km_25}
\end{equation}

\noindent if $\mu \ll 1$. The particles with $E \in (-\Delta
E_\mathrm{cr}, 0)$ move chaotically. Thus, Equation~(\ref{Km_25})
represents the ``$\mu^{2/5}$ law''.

It is interesting that the power-law index in the scaling
$r_\mathrm{H} \propto \mu^{1/3}$ for the radius of the {\it
regular} zone (the Hill sphere, see \citealt{MD99}) around the
secondary is intermediate between the indices in the Wisdom and
``Kepler-map'' scalings, given by Equations~(\ref{em_27}) and
(\ref{Km_25}), respectively. Thus, the indices form a sequence:
2/5, 2/6, 2/7.

Let us characterize the size of the central continuous chaotic
zone in the space of orbital elements. Linearizing the Kepler
map~(\ref{kmp_gm1}) in $y$ near the fixed point at the border of
the map's chaotic layer (thus, the Kepler map is locally
approximated by the standard map, which is a mathematical model of
a multiplet of equally spaced and equally sized resonances, see
\citealt{S14}), one finds for the location of the border:

\begin{equation}
y_\mathrm{cr} = \left( \frac{3 \lambda}{2 K_G} \right)^{2/5} ,
\label{ycr}
\end{equation}

\noindent where $K_G=0.971635406\dots$ \citep{S07IAU}. Using
Equations~(\ref{kmp_Wqasy}) and (\ref{kmp_la}) for $W$ and
$\lambda$, one arrives at

\begin{equation}
\Delta E_\mathrm{cr} = \left| E_\mathrm{cr} \right| = \left| W
y_\mathrm{cr}  \right| \simeq A \mu^{2/5} q^{-1/10} \exp \left(- B
q^{3/2}\right) ,
\label{DE_Km}
\end{equation}

\noindent where

$$
A = 2^{-1/2} 3^{2/5} \pi^{3/5} K_G^{-2/5} = 2.2061\dots, \qquad B
= 2^{5/2} / 15 = 0.3771\dots .
$$

The particle's critical eccentricity $e_\mathrm{cr}$, following
from the relation $\Delta E_\mathrm{cr} = - E_\mathrm{cr} = 1/(2
a_\mathrm{cr}) = (1 - e_\mathrm{cr})/(2 q)$, is

\begin{equation}
e_\mathrm{cr} = 1 - 2 q \Delta E_\mathrm{cr} ,
\label{Km_ecr}
\end{equation}

\noindent where $\Delta E_\mathrm{cr}$ is given by
Equation~(\ref{DE_Km}). The orbits with $e \gtrsim
e_\mathrm{cr}(q)$ are chaotic.

What if $\mu \approx 1/2$, i.e., the binary is approximately
equal-mass? This is quite common in stellar binaries. According to
\cite[formula~(26)]{RH03}, the energy increment in the $\mu = 1/2$
case in the restricted problem limit is given by

\begin{equation}
\delta E \simeq - 2^{7/4} \pi^{1/2} q^{3/4} \exp \left( -
\frac{2^{5/2} q^{3/2}}{3} \right) \sin 2 g_{i} .
\label{dw4}
\end{equation}

\noindent The possibility to use the Kepler map at moderate and
high values of $\mu$ (i.e., at $\mu \sim 1/2$) was discussed in
\cite{S10}. Note that, in Equations~(\ref{kmp_gm2}), the harmonic
term in the first line ($\propto \sin g_{i}$) is just the first
most prominent one in the Fourier expansion of the energy
increment, if $\mu \ll 1$ \citep{P86,S11}. If one increases $\mu$,
the second harmonic ($\propto \sin 2 g_{i}$) becomes more and more
important. If $\mu = 1/2$, the first harmonic ($\propto \sin
g_{i}$) disappears, whereas the second one ($\propto \sin 2
g_{i}$) becomes largest in the series expansion because, due to
the equality of the primaries' masses, the perturbation frequency
is effectively doubled. Thus, the Kepler map, formally, takes the
form

\begin{eqnarray}
E_{i+1} &=& E_{i} + W_{1/2} \sin 2 g_{i} , \nonumber \\
g_{i+1} &=& g_{i} + 2 \pi \vert 2 E_{i+1} \vert^{-3/2} ,
\label{kmp_mu05}
\end{eqnarray}

\noindent where

\begin{equation}
W_{1/2} \simeq - 2^{7/4} \pi^{1/2} q^{3/4} \exp \left( -
\frac{2^{5/2} q^{3/2}}{3} \right) .
\label{Wmu05}
\end{equation}

\noindent By means of substitution $E = W_{1/2} y$, $g = x/2$,
map~(\ref{kmp_mu05}) is reducible to map~(\ref{kmp_gm1}) with

\begin{equation}
\lambda = 2^{1/2} \pi W_{1/2}^{-3/2} ,
\label{kmp_la_mu05}
\end{equation}

\noindent and $y_\mathrm{cr}$ is given by Equation~(\ref{ycr}).
Thus, in the $\mu = 1/2$ case one has

\begin{equation}
\Delta E_\mathrm{cr} = \left| W_{1/2} y_\mathrm{cr}  \right|
\simeq A_{1/2} q^{3/10} \exp \left(- B_{1/2} q^{3/2}\right) ,
\label{DE_mu05}
\end{equation}

\noindent where

$$
A_{1/2} = 2^{1/2} 3^{2/5} \pi^{3/5} K_G^{-2/5} = 4.4122\dots,
\qquad B_{1/2} = 2^{7/2} / 15 = 0.7542\dots .
$$

\noindent The critical eccentricity $e_\mathrm{cr}$ is given by
formula~(\ref{Km_ecr}).

As illustrated in the next section, the critical curve given by
Equation~(\ref{DE_mu05}) looks somewhat different from a curve
resulting from convergence of the critical curves given by
Equation~(\ref{DE_Km}) at $\mu \to 1/2$, though the locations of
both curves are approximately the same. The difference is due to
the fact that the actual energy increment is not given by a single
harmonic term, but is a Fourier series of harmonic terms
\citep{P86,LS94}, from which we have taken only the leading terms:
the term $\propto \sin g_{i}$ in the case of $\mu \ll 1$, and the
term $\propto \sin 2 g_{i}$ in the case of $\mu \approx 1/2$.
Taking into account additional harmonic terms is important, in
particular, at $q$ close to~1 (see \citealt{LS94}).

\section{Size of the circumbinary chaotic zone}
\label{sec_size}

\begin{figure}[h!]
\begin{center}
\includegraphics[width=12cm]{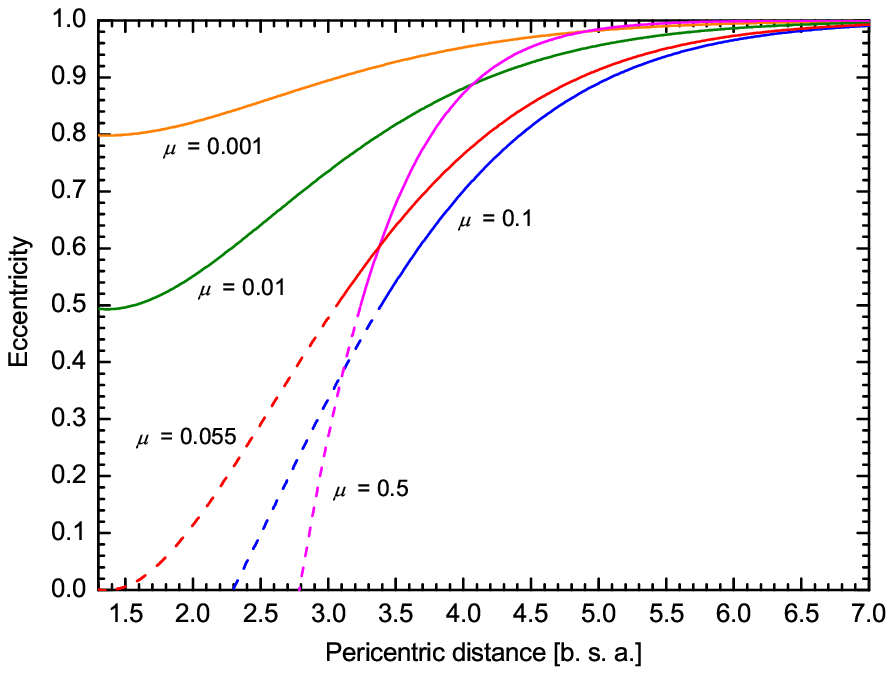}
\end{center}
\caption{Critical curves, given by
Equation~(\protect\ref{Km_ecr}), for several values of the mass
parameter $\mu$. The curves separate chaotic and regular domains
(chaos is on the left). Extrapolations are dashed. The pericentric
distance is measured in the units of the central binary's
semimajor axis (abbreviated as ``b.~s.~a.'' at the horizontal axis
caption). } \label{fig1}
\end{figure}

In Figure~\ref{fig1}, theoretical dependences ``pericentric
distance $q$--critical eccentricity $e_\mathrm{cr}$'', are
constructed for several values of $\mu$, using the formulae
derived above. Global chaos extends to the left of the curves.
Tentatively extrapolating the critical curves to zero
eccentricity, one sees that, if $\mu=0.1$, the curve hits the
horizontal axis at $q=2.3$, and, from Kepler's third law, this
corresponds to the ratio $\approx 3.5$ between the orbital periods
of the particle and the binary. The value of $q=2.8$ (where the
critical curve hits the horizontal axis at $\mu=0.5$) corresponds
to the ratio $\approx 4.7$. Thus, the extrapolation of the
critical curve to zero eccentricity gives $q$ rather insensitive
to $\mu$ in the range $\mu \sim 0.1$--$0.5$, which is typical for
binary stars. If $\mu$ is in this range, the chaotic zone boundary
lies in the region of resonances from 7/2 to 5/1.

We extrapolate the curves given by the Kepler map theory to the
tertiary's low eccentricities. To emphasize the extrapolative
character of the curves at low eccentricities, they are dashed at
$e < 0.5$. However, the theory was developed for high
eccentricities. The possibility of such an extrapolation is
justified post factum: the curves corresponding to $\mu \sim
0.1$--$0.5$ hit the $e=0$ axis at high enough values of $q$, at
which the higher-order harmonics in the Fourier expansion of the
energy increment are relatively unimportant because these
harmonics are exponentially small with the harmonic order $j$
(they are proportional to $\exp (-4j q^{3/2} / (2^{1/2} 3))$;
\citealt{P86,PB88}).

Thus, according to the theoretical dependences presented in
Figure~\ref{fig1}, the central chaotic zone's radial size,
measured in the units of binary's semimajor axis, is $\sim 3$ at
moderate eccentricities, and $\sim 2.3$--2.8 at zero
eccentricities of the tertiaries, if one considers primaries of
comparable masses ($\mu \sim 0.1$--0.5).

At $e=0$, the estimate can be compared to that given by the
numerical-experimental criterion of~\cite{HW99}. Since
fit~(\ref{HWcrit}) was accomplished in \cite{HW99} for $\mu \geq
0.1$, we make comparisons at $\mu \geq 0.1$. Setting
$e_\mathrm{b}=0$ in Equation~(\ref{HWcrit}), for $\mu = 0.1$ and
$0.5$ one obtains $a_\mathrm{cr}/a_\mathrm{b} = 2.0$ and $2.4$,
whereas Figure~\ref{fig1} gives $a_\mathrm{cr}/a_\mathrm{b} = 2.3$
and $2.8$, respectively. The agreement, taking into account the
extrapolative character of our predictions and the strongly
``ragged'' character of the global chaos border at low values of
$e$ (see Section~\ref{sec_diag}), can be considered rather good,
though a systematic shift is present.

Our results at $e=0$ can be also compared to semianalytical data
presented in \cite{Sz80} and \cite{SzM81}, who employed
computations of the topology of the zero velocity curves in the
circular restricted three-body problem in the framework of the
Hill--Lyapunov approach. At $\mu = 0.1$, $0.24$, and $0.5$,
\cite{Sz80} and \cite{SzM81} obtained $a_\mathrm{cr}/a_\mathrm{b}
\approx 2.24$, $2.4$, and $2.17$, whereas, from our formulas, one
has $a_\mathrm{cr}/a_\mathrm{b} \approx 2.29$, $2.91$, and $2.79$,
respectively. At $\mu = 0.1$ the agreement is perfect, and it is
even better than that with the data of \cite{HW99}, but at $\mu =
0.5$ the divergence is rather large. The nature of this divergence
needs further analysis. A comparison of the results of
\cite{SzM81} with fit~(\ref{HWcrit}) is discussed in~\cite{HW99}.

\section{The mass parameter threshold}
\label{sec_thr}

In Figure~\ref{fig1}, the theoretical curves start (on increasing
$\mu$) to hit the horizontal axis $e=0$ at $\mu \approx 0.0547
\sim 0.05$; thus, this value of $\mu$ can be considered as an
approximate threshold value at which the central continuous
chaotic zone, emerging due to the overlap of the $p$:1 resonances,
appears at all eccentricities of the tertiaries. (Note that our
analysis solely concerns the circumbinary orbits; for the
circumcomponent (satellite-type) orbits, stable orbits always
exist inside the Hill spheres of the binary components.)

This threshold has a notable physical meaning: above it, the
tertiary, even starting from a small eccentricity, can diffuse,
following the sequence of the overlapping $p$:1 resonances, up to
ejection from the system; close encounters with other bodies are
not required for the escape. Below it, the diffusion in the
overlapping $(p+1)$:$p$ resonances does not lead, in itself, to
the ejection; the chaotic band surrounding the secondary's orbit
is ``cleaned up'' due to close encounters of the tertiaries with
the secondary.

The prediction for the threshold relies on the mentioned rather
sharp transition in the behavior of the extrapolated critical
curves, taking place at $\mu \sim 0.05$. The actual threshold
$\mu$ value can differ somewhat from the extrapolative one. Is
there any independent evidence on the threshold $\mu$?

First of all, let us note that the threshold existence does not
imply that, at $\mu$ less than the threshold, any initial circular
orbit external to the binary is stable, regardless of the
semimajor axis. Indeed, below the threshold, Wisdom's chaotic band
(emerging due to overlapping of $p$:1 resonances; see
Section~\ref{sec_crit}) around the orbit of the perturber is
``unveiled''. On the other hand, if one extrapolates the
polynomial fit~(\ref{HWcrit}) to zero $\mu$, in the circular
problem ($e_\mathrm{b} = 0$) one gets $a_\mathrm{cr}/a_\mathrm{b}
= 1.6$, whereas in reality in this limit there is no chaos at all,
and the width of Wisdom's chaotic band is zero, as follows from
Equations~(\ref{em_27}) or~(\ref{actem_27}). Thus, a notable
transition between the Holman--Wiegert and Wisdom relations may
take place, somewhere in the interval $0.001 \lesssim \mu \lesssim
0.1$. The interval is such because Wisdom's law was verified in
numerical experiments at least up to $\mu = 0.001$
\citep{MD99,QF06}, and \cite{HW99} obtained fit~(\ref{HWcrit}) at
$\mu \geq 0.1$.

\begin{figure}[h!]
\begin{center}
\includegraphics[width=12cm]{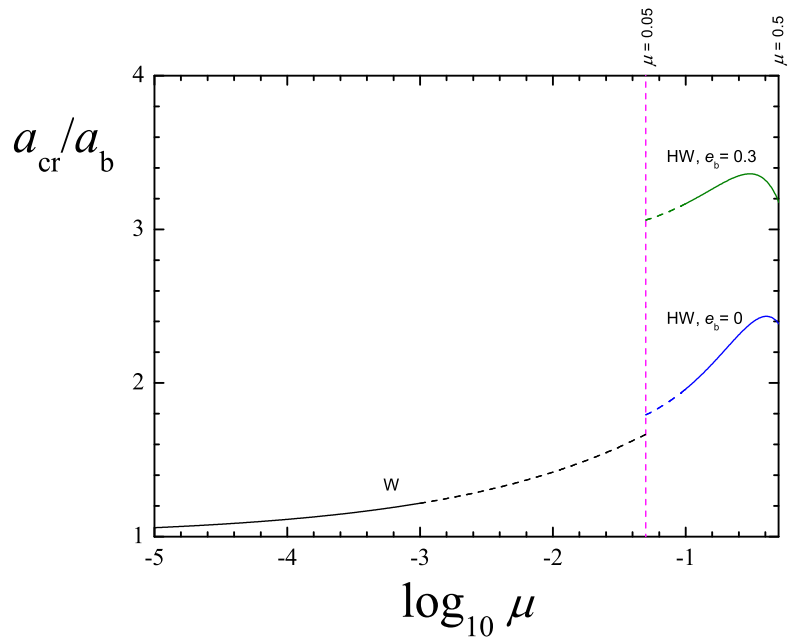}
\end{center}
\caption{``Mass parameter--critical semimajor axis'' analytical
relationships. Curve ``W'': Wisdom's law, given by
Equation~(\protect\ref{actem_27}). Curves ``HW'':
Holman--Wiegert's empirical relation, given by
Equation~(\protect\ref{HWcrit}), at two values of the central
binary eccentricity. Extrapolations are dashed. The vertical
dashed (magenta) line shows the theoretical threshold $\mu =
0.05$. } \label{fig2}
\end{figure}

Let us look at the ``junction'' of the two relations in more
detail. In Figure~\ref{fig2}, the ``mass parameter--critical
semimajor axis'' analytical relationships are presented
graphically. Wisdom's law is valid, as derived, at $e_\mathrm{b} =
0$; Holman--Wiegert's curve, constructed for $e_\mathrm{b} = 0$,
joins Wisdom's curve, as one can see, rather smoothly, though a
moderate jump may be present in the ``uncertainty interval''
around the $\mu$ threshold value.

As pointed out in \cite{QF06}, at $\mu \leq 0.001$ and
$e_\mathrm{b} < 0.3$ the chaotic zone size is independent of
$e_\mathrm{b}$ and is described by Wisdom's law. Therefore, it is
adequate to compare how Wisdom's and Holman--Wiegert's relations
join, if $e_\mathrm{b} = 0.3$. From Figure~\ref{fig2}, it is
evident that a jump or a sharp rise should be definitely present
in the ``uncertainty interval'', i.e., the chaotic zone size
increases sharply somewhere in the interval. In other words, in
the eccentric case, the transition from one mechanism of chaos
generation (overlap of $(p+1)$:$p$ resonances) to another one
(overlap of $p$:1 resonances) seems to result not only in the
change of the diffusion character, but also in the sharp increase
of the chaotic zone size.

Independent evidence for the $\mu$ threshold value follows from
the fact that the derived $\mu$ threshold roughly corresponds to
the $\mu$ value at which the loss of stability of the triangular
Lagrangian points L$_4$ and L$_5$ takes place (this value is
$\approx 0.04$, see \citealt{Sz67}). Though this might seem to be
merely a coincidence, physically it looks quite natural that the
transition to global chaos (due to overlap of $p$:1 resonances)
leaves no place for regular islands in the phase space around the
triangular libration points.

Besides, the threshold existence seems to explain an old
numerical-experi\-mental result by \cite{N76} that the
Sun--Jupiter--Saturn system becomes unstable on increasing $\mu$
29 times, i.e., up to $\sim 0.03$---rather close to $0.05$, taking
into account that the problem differs from the restricted one.
Recently, \cite{KhK11} obtained an even smaller
numerical-experimental value of $\mu$ for the system gross
instability upsurge, namely, $\mu \sim 0.02$. Again, since the
problem setting is somewhat different, one may say that the
numerical-experimental value roughly agrees with our theoretical
prediction. The differences between the theoretical and
numerical-experimental values can be due to the extrapolative
character of the former as well.

Further on, in Section~\ref{sec_thr_obs}, we compare the
theoretical prediction for the threshold directly with relevant
observational data on exoplanetary systems.

\section{Stability diagrams and the criterion predictions}
\label{sec_diag}

Let us consider an example of our theory application concerning
circumbinary exoplanets. Nowadays, several circumbinary planets in
systems of main-sequence binaries are known; Kepler-16b is the
prototype, discovered in 2011 by \cite{D11}. In \cite{PS13},
stability diagrams in the ``pericentric distance--eccentricity''
plane were constructed, which show that Kepler-16b is in a
hazardous vicinity to the global instability domain: the planet
resides just between the instability ``teeth'' in the space of
orbital parameters. Kepler-16b is safe inside a resonance cell
bounded by the unstable 5/1 and 6/1 resonances. The planets
Kepler-34b and Kepler-35b, reported in \cite{W12}, are also safe
inside resonance cells at the chaos border \citep{PS13}.

\begin{figure}[h!]
\begin{center}
\includegraphics[width=12cm]{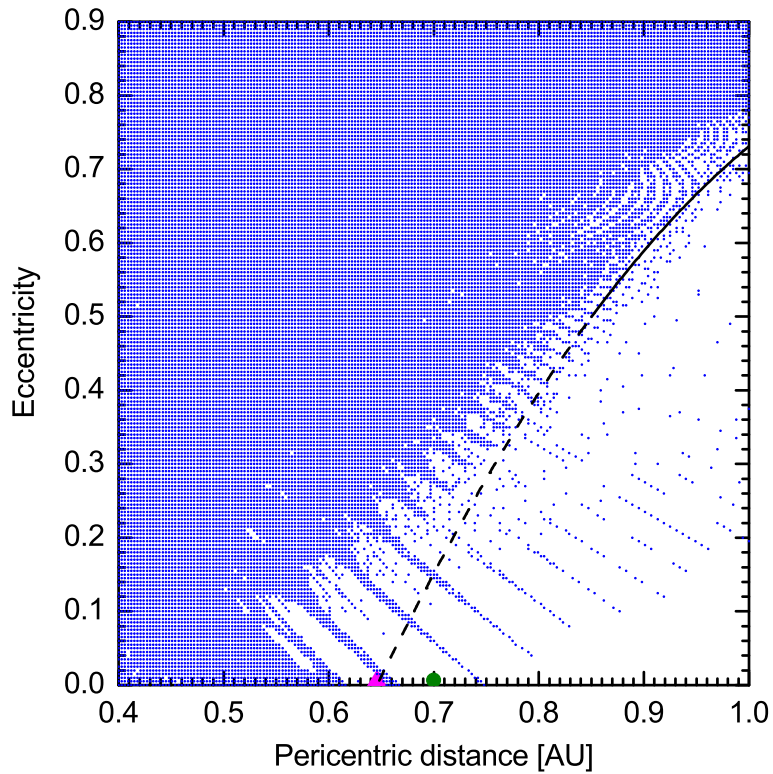}
\end{center}
\caption{Stability diagram for Kepler-16b. The chaotic domain is
shown by the shaded area, as revealed numerically in
\protect\cite{PS13}. The critical curve, given by
Equation~(\protect\ref{Km_ecr}) at $\mu = 0.227$, is superimposed
(extrapolation is dashed). The green dot indicates the actual
location of planet Kepler-16b. The red triangle indicates the
instability border location, as given by the Holman--Wiegert
criterion at zero eccentricity. } \label{fig3}
\end{figure}

In Figure~\ref{fig3}, the curve $e_\mathrm{cr}(q)$, given by
Equation~(\ref{Km_ecr}), is superimposed on the stability diagram
constructed in \cite{PS13} for Kepler-16b by means of numerically
integrating the equations of planetary motion and computing the
Lyapunov spectra. The motion is regarded as chaotic, if the
maximum Lyapunov exponent is non-zero. The location of planet
Kepler-16b is shown in Figure~\ref{fig3} by a green dot. The
instability border location, as given by the Holman--Wiegert
criterion (Equation~(\ref{HWcrit})), is shown by a red triangle;
note that its location coincides with the extrapolation of the
curve $e_\mathrm{cr}(q)$ to zero eccentricity. (Such an agreement
is in fact rather fortuitous because the Holman--Wiegert value has
been calculated here for the actual value of the binary
eccentricity $e_\mathrm{b}=0.159$, whereas the analytical
expression for $e_\mathrm{cr}(q)$ was derived setting
$e_\mathrm{b}=0$.)

One can see that the $e_\mathrm{cr}(q)$ curve approximately
describes the smoothed border of the chaotic zone in
Figure~\ref{fig3}. In fact, the real border is ragged; the most
prominent ``teeth'' correspond to the integer $p$:1 resonances.
The Farey tree \citep{M92} of resonant ``teeth'' at the border is
evident. (Consider the lowest order ``neighboring'' resonances
$m/n$ and $m'/n'$; in the given case, these are the integer mean
motion resonances $m/1$ and $(m+1)/1$. The lower level of the
Farey tree is made of ``mediants'' given by the formula $m''/n''=
(m + m')/(n + n') = (2m+1)/2$. Thus, the half-integer mean motion
resonances are the mediants for the integer ones, and so on.) The
resonances densely accumulate higher in the diagram, on
approaching the parabolic separatrix. The diagram graphically
demonstrates how the resonances (beginning with the 4/1, 5/1, 6/1
resonances on the left of the figure) overlap. Note that the 5/1
nd 6/1 resonant ``teeth'' engulf the cell where the planet is
located.

\section{The mass parameter threshold and the diversity of observed exosystems}
\label{sec_thr_obs}

In this section, we discuss the relevance of the mass parameter
threshold to the diversity of the observed orbital configurations
of exosystems. How well does the theoretical prediction for the
$\mu$ threshold agree with the observational data? To explore
this, let us construct an empirical relationship between the
primaries' mass parameter $\mu$ and the ratio of the tertiary and
secondary orbital periods $T_\mathrm{out}/T_\mathrm{in}$, based on
a relevant sample of exosystems. For this purpose, we use the
exoplanetary data provided by the Exoplanet Encyclopedia
(www.exoplanet.eu), as on 2014 June 30.

Two classes of exosystems are directly relevant: biplanetary (a
star plus two planets) and circumbinary (two stars plus one planet
orbiting them both). Besides, for our criterion to be applicable,
it is required that the planet in the outermost orbit have the
smallest mass in the system. Solely, such systems have been
included in the sample.

\begin{figure}[h!]
\begin{center}
\includegraphics[width=12cm]{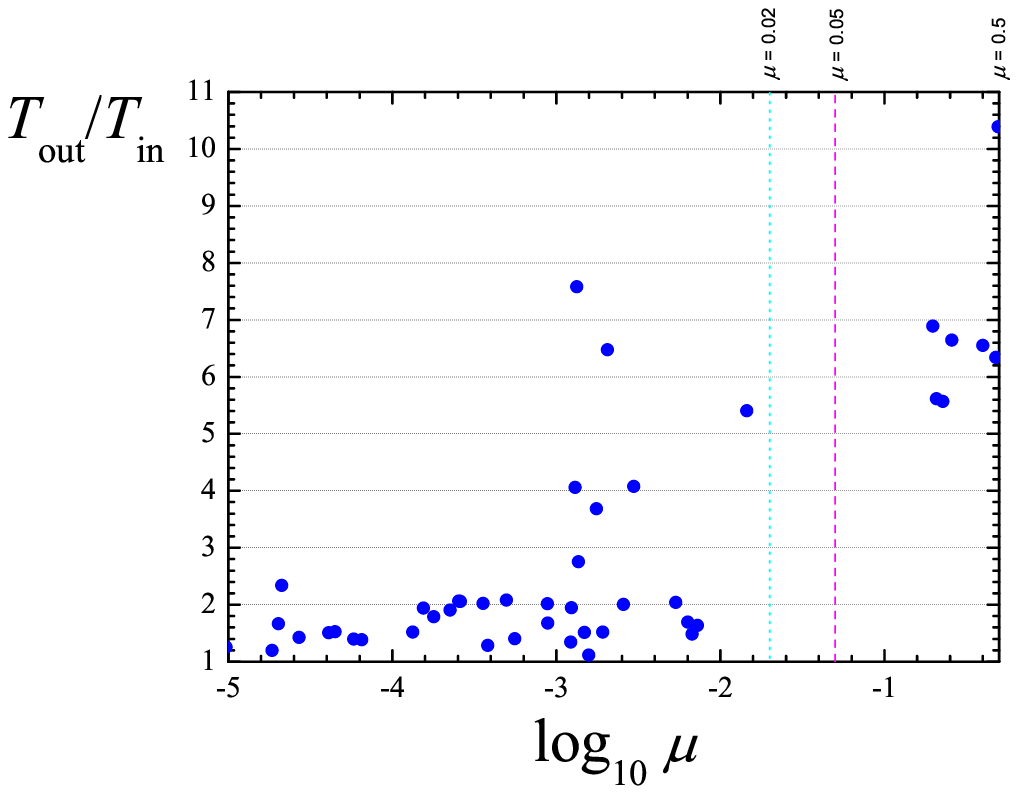}
\end{center}
\caption{``Mass parameter--orbital period ratio'' relationship for
the biplanetary and circumbinary exosystems (dots). The vertical
dashed (magenta) line shows the theoretical threshold $\mu =
0.05$. The dotted (cyan) line corresponds to $\mu = 0.02$, see the
text. } \label{fig4}
\end{figure}

The resulting plot is shown in Figure~\ref{fig4}. The dots show
the location of the exosystems. The biplanetary systems all turn
out to be on the left of two vertical (dotted and dashed) lines,
whereas the circumbinary systems are on the right of them. The
vertical dashed (magenta) line indicates the theoretical threshold
$\mu = 0.05$ for the appearance of the central chaotic zone. The
dotted (cyan) line is drawn at $\mu = 0.02$, roughly corresponding
to the numerical-experimental results by \cite{N76} and
\cite{KhK11} on modeling the upsurge of instability of the
Sun--Jupiter--Saturn system on raising the system mass parameter.

In Figure~\ref{fig4}, the total absence of exosystems with
$T_\mathrm{out}/T_\mathrm{in} < 5$ at $\mu > 0.05$ is evident, in
agreement with our theoretical prediction that, at $\mu > 0.05$,
the central chaotic zone is formed, where the particle orbits with
any initial eccentricities are subject to the unlimited chaotic
diffusion, up to ejection from the system.

Two comments are in order. (1)~A certain gap in the $\mu$ values
exists between the two classes of exosystems comprising the
sample. Indeed, in the biplanetary case the mass ratio of a
central star--planet binary of the ``Solar--Jovian'' type is $\sim
0.001$, whereas in the circumbinary case the mass ratio of the
typical main-sequence binary is $\sim 1$. We believe that the gap
will be filled in the future, when more exosystems that include
brown dwarfs are observed, such as systems with the primaries
composed by a main-sequence star plus a brown dwarf, or systems
with primaries composed by a brown dwarf plus a Jovian-type
planet. A prototype of the former class system is HD202206 ($\mu
\approx 0.014$), in which the inner ``planet'' with a mass of
$\approx 17.4$ Jovian masses is most likely a brown dwarf
\citep{C05}; in Figure~\ref{fig4}, the dot corresponding to this
system is closest to the dotted (cyan) line. (2)~As follows from
Figure~\ref{fig4}, many exosystems with $\mu < 0.01$ cluster at
the 2/1 orbital resonance; whereas the exosystems with $\mu > 0.1$
do not seem to cluster at any integer resonance, but rather at
half-integer ones. The latter fact is in accord with finding by
\cite{PS13} that the observed circumbinary planets survive (though
located close to the global instability border in the space of
orbital elements) because they are safe inside resonance cells
formed by unstable high-order integer resonances.

\section{Conclusions}
\label{sec_concl}

Our main conclusions are as follows.

\noindent 1.~The presence of the continuous chaotic zone around
the gravitating binary of comparable masses is explained as being
due to the overlap of resonances $p$:1 between the tertiary and
the central binary, overlapping already at moderate values of $p
\sim 4$--5 and accumulating at $p \to \infty$ near the parabolic
separatrix.

\noindent 2.~At the tertiaries' moderate eccentricities, the size
of the continuous chaotic zone is about thrice the size of the
binary, and it increases with the tertiary eccentricity.

\noindent 3.~The binary's mass ratio, above which such a zone is
present, is estimated to be $\sim 0.05$. This threshold has a
notable physical meaning: above it, the tertiary, even starting
from a small eccentricity, can diffuse, following the sequence of
the overlapping $p$:1 resonances, up to ejection from the system;
close encounters with other bodies are not required for escape.

\noindent 4.~The diversity of the observed orbital configurations
of biplanetary and circumbinary exosystems is in accord with the
existence of the primaries' mass parameter threshold at $\mu \sim
0.02$--$0.05$.

\bigskip

\noindent {\bf \Large Acknowledgments}

\medskip

The author is grateful to the referee for useful remarks. The
author wish to thank T.V.~Demidova, A.V.~Melnikov, V.V.~Orlov and
E.A.~Popova for helpful discussions and comments. This work was
supported in part by the Russian Foundation for Basic Research
(projects Nos.\ 12-02-00185 and 14-02-00464) and by the Programmes
of Fundamental Research of the Russian Academy of Sciences
``Fundamental Problems in Nonlinear Dynamics'' and ``Fundamental
Problems of the Solar System Studies and Exploration''.

\newpage

\begin{thebibliography}{}

\bibitem[Artymowicz \& Lubow(1994)]{AL94}
Artymowicz, P., \& Lubow, S. H. 1994, ApJ, 421, 651

\bibitem[Artymowicz \& Lubow(1996)]{AL96}
Artymowicz, P., \& Lubow, S. H. 1996, in Disks and Outflows around
Young Stars, ed. S.~V.~W.~Beckwith, J.~Staude, A.~M.~Quetz, \&
A.~Natta (New York: Springer), 115

\bibitem[Chirikov(1959)]{C59}
Chirikov, B. V. 1959, Atomnaya Energiya, 6, 630 (1960, J. Nucl.
Energy Part C: Plasma Phys., 1, 253)

\bibitem[Chirikov(1979)]{C79}
Chirikov, B. V. 1979, PhR, 52, 263

\bibitem[Chirikov \& Vecheslavov(1986)]{CV86}
Chirikov, B. V., \& Vecheslavov, V. V. 1986, INP Preprint 86--184
(Novosibirsk: Institute of Nuclear Physics),
http://www.quantware.ups-tlse.fr/chirikov/ publbinp.html

\bibitem[Chirikov \& Vecheslavov(2000)]{CV00}
Chirikov, B. V., \& Vecheslavov, V. V. 2000, J. Exp. Theor. Phys.,
90, 562 (Zh.\ Eksp.\ Teor.\ Fiz., 117, 644)

\bibitem[Correia et al.(2005)]{C05}
Correia, A. C. M., Udry, S., Mayor, M., et al. 2005, A\&A, 440,
751

\bibitem[Doyle et al.(2011)]{D11}
Doyle, L. R., Carter, J. A., Fabrycky, D. C., et al. 2011,
Science, 333, 1602

\bibitem[Duncan et al.(1989)]{DQT89}
Duncan, M., Quinn, T., \& Tremaine, S. 1989, Icarus, 82, 402

\bibitem[Holman \& Wiegert(1999)]{HW99}
Holman, M. J., \& Wiegert, P. A. 1999, AJ, 117, 621

\bibitem[Kholshevnikov \& Kuznetsov(2011)]{KhK11}
Kholshevnikov, K. V., \& Kuznetsov, E. D. 2011, Celest. Mech. Dyn.
Astron., 109, 201

\bibitem[Lichtenberg \& Lieberman(1992)]{LL92}
Lichtenberg, A. J., \& Lieberman, M. A. 1992, Regular and Chaotic
Dynamics (New York: Springer)

\bibitem[Liu \& Sun(1994)]{LS94}
Liu, J., \& Sun, Y. S. 1994, Celest. Mech. Dyn. Astron., 60, 3

\bibitem[Mardling(2008)]{Ma08}
Mardling, R. A. 2008, in The Cambridge N-Body Lectures (Lecture
Notes in Physics, Vol.~760; Berlin: Springer), 59

\bibitem[Meiss(1992)]{M92}
Meiss, J. D. 1992, Rev. Mod. Phys., 64, 795

\bibitem[Meschiari(2012)]{M12}
Meschiari, S. 2012, ApJ, 752, 71

\bibitem[Mikkola(2008)]{M08}
Mikkola, S. 2008, in Multiple Stars Across the H-R Diagram, ed.\
S.~Hubrig, M.~Petr-Gotzens, \& A.~Tokovinin (ESA Astrophysics
Symposia; Berlin: Springer), 11

\bibitem[Moriwaki \& Nakagawa(2004)]{MN04}
Moriwaki, K., \& Nakagawa, Y. 2004, ApJ, 609, 1065

\bibitem[Mudryk \& Wu(2006)]{MW06}
Mudryk, L. R., \& Wu, Y. 2006, ApJ, 639, 423

\bibitem[Murray \& Dermott(1999)]{MD99}
Murray, C. D., \& Dermott, S. F. 1999, Solar System Dynamics
(Cambridge: Cambridge Univ. Press)

\bibitem[Nacozy(1976)]{N76}
Nacozy, P. E. 1976, AJ, 81, 787

\bibitem[Paardekooper et al.(2012)]{PLTB12}
Paardekooper, S.-J., Leinhardt, Z. M., Th\'ebault, T., et al.
2012, ApJL, 754, L16

\bibitem[Petrosky(1986)]{P86}
Petrosky, T. Y. 1986, Phys. Lett. A, 117, 328

\bibitem[Petrosky \& Broucke(1988)]{PB88}
Petrosky, T. Y., \& Broucke, R. 1988, Celest. Mech. Dyn. Astron.,
42, 53

\bibitem[Pierens \& Nelson(2007)]{PN07}
Pierens, A., \& Nelson, R. P. 2007, A\&A, 472, 993

\bibitem[Pierens \& Nelson(2008)]{PN08}
Pierens, A., \& Nelson, R. P. 2008, A\&A, 483, 633

\bibitem[Popova \& Shevchenko(2012)]{PS12}
Popova, E. A., \& Shevchenko, I. I. 2012, Astron.\ Lett., 38, 581
(Pis'ma Astron.\ Zhurnal, 38, 652)

\bibitem[Popova \& Shevchenko(2013)]{PS13}
Popova, E. A., \& Shevchenko, I. I. 2013, ApJ, 769, 152

\bibitem[Quillen \& Faber(2006)]{QF06}
Quillen, A. C., \& Faber, P. 2006, MNRAS, 373, 1245

\bibitem[Roy \& Haddow(2003)]{RH03}
Roy, A., \& Haddow, M. 2003, Celest. Mech. Dyn. Astron., 87, 411

\bibitem[Saito et al.(2012)]{STO12}
Saito, M. M., Tanikawa, K., \& Orlov, V. V. 2012, Celest. Mech.
Dyn. Astron., 112, 235

\bibitem[Saito et al.(2013)]{STO13}
Saito, M. M., Tanikawa, K., \& Orlov, V. V. 2013, Celest. Mech.
Dyn. Astron., 116, 1

\bibitem[Shevchenko(2007)]{S07IAU}
Shevchenko, I. I. 2007, in IAU Symp. 236, Near Earth Objects, Our
Celestial Neighbors: Opportunity and Risk, ed. A.~Milani,
G.~B.~Valsecchi, \& D.~Vokrouhlick\'y (Cambridge: Cambridge Univ.
Press), 15

\bibitem[Shevchenko(2008)]{S08}
Shevchenko, I. I. 2008, MNRAS, 384, 1211; 2010, MNRAS, 407, 704

\bibitem[Shevchenko(2010)]{S10}
Shevchenko, I. I. 2010, Phys. Rev. E, 81, 066216

\bibitem[Shevchenko(2011)]{S11}
Shevchenko, I. I. 2011, New Astronomy, 16, 94

\bibitem[Shevchenko(2014)]{S14}
Shevchenko, I. I. 2014, Phys. Lett. A, 378, 34

\bibitem[Szebehely(1967)]{Sz67}
Szebehely, V. 1967, Theory of Orbits (New York: Academic Press)

\bibitem[Szebehely(1980)]{Sz80}
Szebehely, V. 1980, Celest. Mech., 22, 7

\bibitem[Szebehely \& McKenzie(1981)]{SzM81}
Szebehely, V., \& McKenzie, R. 1981, Celest. Mech., 23, 3

\bibitem[Valtonen et al.(2008)]{V08}
Valtonen, M., Myll\"ari, A., Orlov, V., et al. 2008, in IAU Symp.
246, Dynamical Evolution of Dense Stellar Systems, ed.
E.~Vesperini, M.~Giersz, \& A.~Sills (Cambridge: Cambridge Univ.
Press), 209

\bibitem[Welsh et al.(2012)]{W12}
Welsh, W. F., Orosz, J. A., Carter, J. A., et al. 2012, Nature,
481, 475

\bibitem[Welsh et al.(2014)]{W14}
Welsh, W. F., Orosz, J. A., Carter, J. A., et al. 2014, in IAU
Symp. 293, Formation, Detection, and Characterization of
Extrasolar Habitable Planets, ed. N.~Haghighipour (Cambridge:
Cambridge Univ. Press), 125

\bibitem[Wisdom(1980)]{W80}
Wisdom, J. 1980, AJ, 85, 1122

\end{thebibliography}
\end{document}